\documentclass[%
reprint,
 amsmath,amssymb,
prx,
floatfix
]{revtex4-2}

\usepackage{graphicx}
\usepackage{dcolumn}
\usepackage{bm}
\usepackage{braket}
\usepackage{mathtools}
\usepackage{tikz}
\usepackage{layouts}
\usetikzlibrary{arrows.meta}
\usetikzlibrary{shapes}
\usetikzlibrary{3d}

\usepackage{newfloat}
\DeclareFloatingEnvironment[fileext=algf,placement={!tbp},name=Algorithm]{algorithm}

\usepackage{algpseudocode}
\usepackage{placeins}
\usepackage[normalem]{ulem}
\usepackage{xcolor}
\usepackage{enumitem}

\newcommand{\changed}[2]{\textcolor{blue}{#1}}
\makeatletter
\newcommand\thefontsize[1]{{#1 The current font size is: \f@size pt\par}}
\makeatother
\begin{document}

\preprint{PRX/Quantum}

\title{Quantum Circuit Optimization using \\ Differentiable Programming of Tensor Network States}

\author{David Rogerson}
  \email{david.rogerson@rutgers.edu}
\author{Ananda Roy}%
  \email{ananda.roy@physics.rutgers.edu}
\affiliation{%
Department of Physics and Astronomy, Rutgers University, Piscataway, NJ 08854-8019 USA
}%
\date{\today}

\begin{abstract}
Efficient quantum circuit optimization schemes are central to  quantum simulation of strongly interacting quantum many body systems. Here, we present an optimization algorithm which combines machine learning techniques and tensor network methods. The said algorithm runs on classical hardware and finds shallow, accurate quantum circuits by minimizing scalar cost functions. The gradients relevant for the optimization process are computed using the reverse mode automatic differentiation technique implemented on top of the time-evolved block decimation algorithm for matrix product states. A variation of the ADAM optimizer is utilized to perform a gradient descent on the manifolds of charge conserving unitary operators to find the optimal quantum circuit. The efficacy of this approach is demonstrated by finding the ground states of spin chain Hamiltonians for the Ising, three-state Potts and the massive Schwinger models for system sizes up to $L=100$. The first ten excited states of these models are also obtained for system sizes $L=24$. All circuits achieve high state fidelities within reasonable CPU time and modest memory requirements.

\end{abstract}
\maketitle
\pagebreak
\begin{figure*}
    \centering
    \begin{tikzpicture}
        \node[anchor=south west,inner sep=0] (image) at (0,0) {\includegraphics[width=0.9\columnwidth]{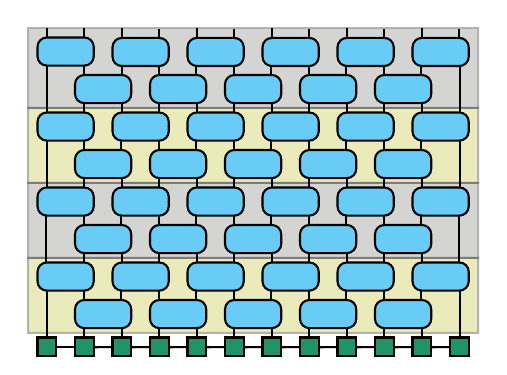}};
        \begin{scope}[x={(image.south east)},y={(image.north west)}, >={Stealth}]
            \node at (0.02, 0.98) {\textbf{(a)}};
            \node at (0.98, 0.07) {$\ket{\psi_i}$};
            \node at (0.5, 0.98) {$\ket{\psi_f}$};
            \node[rotate=90] at (0.0, 0.2) {$\mathcal{D}=1$};
            \node[rotate=90] at (0.0, 0.4) {$2$};
            \node[rotate=90] at (0.0, 0.6) {$3$};
            \node[rotate=90] at (0.0, 0.8) {$4$};
        \end{scope}
    \end{tikzpicture}
    \begin{tikzpicture}
        \node[anchor=south west,inner sep=0] (image) at (0,0) {\includegraphics[width=0.9\columnwidth]{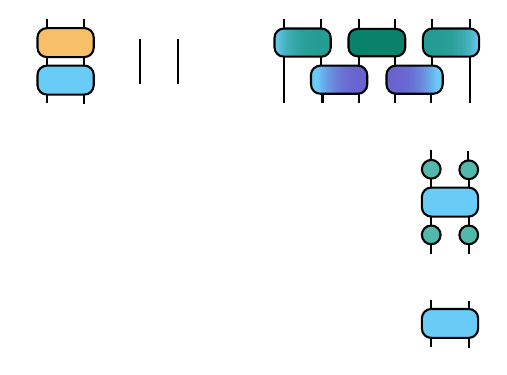}};
        \begin{scope}[x={(image.south east)},y={(image.north west)}, >={Stealth}]
            \node at (0.025, 0.95) {\textbf{(b)}};
            \node at (0.50, 0.95) {\textbf{(c)}};
            \node at (0.77, 0.55) {\textbf{(e)}};
            \node at (0.025, 0.55) {\textbf{(d)}};
            \node at (0.225, 0.84) {$=$};
            \node at (0.135, 0.89) {$U^\dagger$};
            \node at (0.13, 0.79) {$U$};
            \node at (0.85, 0.55) {\tiny{$Z$}};
            \node at (0.925, 0.55) {\tiny{$Z$}};
            \node at (0.85, 0.374) {\tiny{$Z$}};
            \node at (0.925, 0.374) {\tiny{$Z$}};
            \node at (0.89, 0.25) {$=$};
        \end{scope}
            \node[anchor=south west,inner sep=0] (manifold) at (0,0) {\includegraphics[width=0.65\columnwidth]{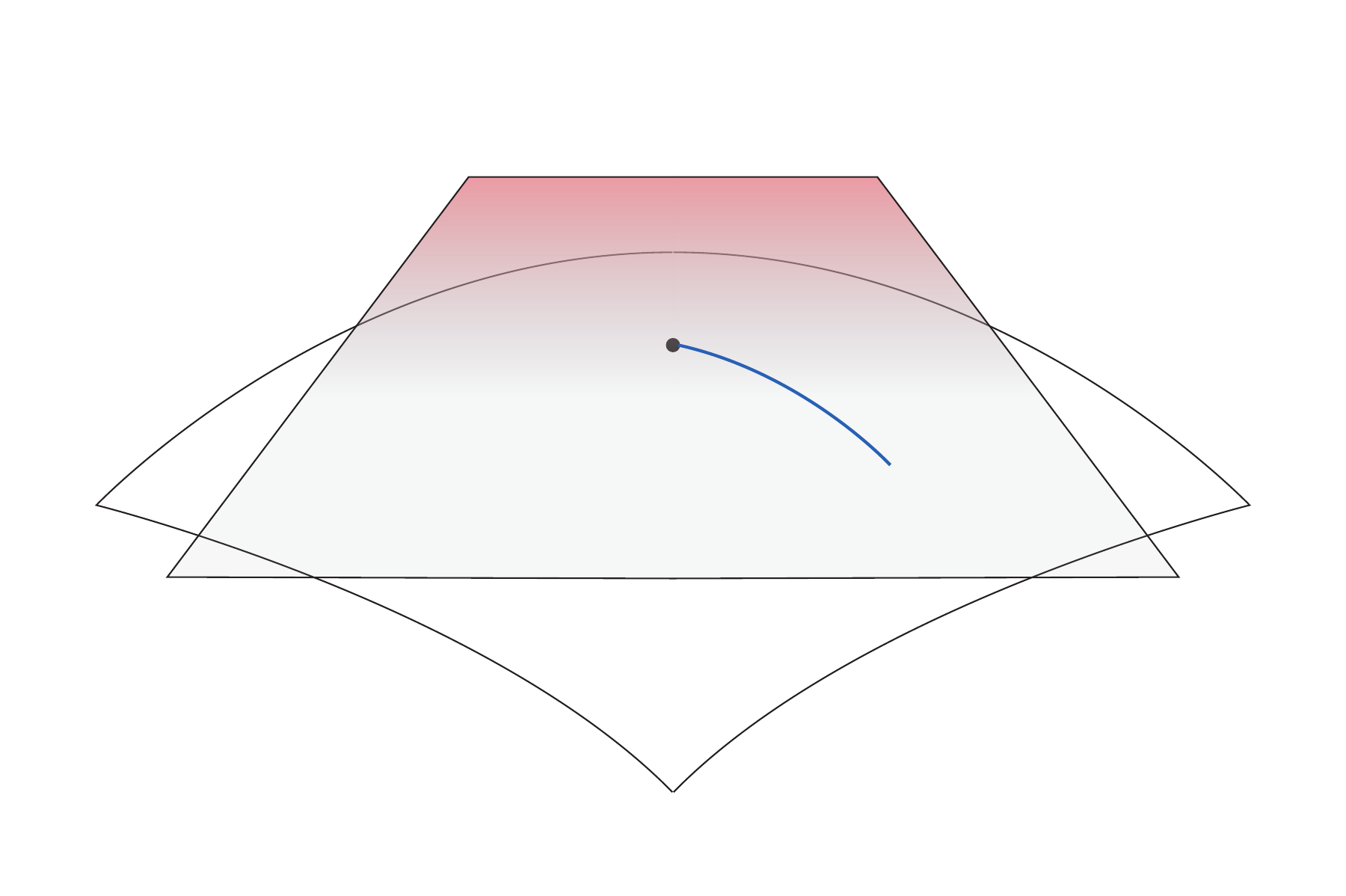}};
            \begin{scope}[x={(manifold.south east)},y={(manifold.north west)}]
                \coordinate (p) at (0.5, 5/8-0.01);
                \coordinate (o) at (0.3, 0.2);
                \draw[->,>={Stealth}] (p) to (0.67,0.57);
                \draw[->, >={Stealth}, color=gray] (p) to  (0.67, 0.9);
                \draw[<-, dashed, >={Stealth},color=gray] (0.67,0.60) to (0.67, 0.87);
                \coordinate (p') at (0.65, 0.47);
                \node[color=gray] () at (0.57, 0.83) {$\nabla$};
                \node[color=blue] (geo) at (0.6,0.5) {$\eta$};
                \node () at (0.28, 0.8) {\footnotesize $T_U M$};
                \node[color=gray] () at (0.83, 0.73) {$\mathrm{lift}(\nabla, U)$};
                \node () at (0.6, 0.65) {$\nabla^R$};
                \node () at (0.5, 5/8-0.07) {$U$};
                \node () at (0.68, 0.44) {$U'$};
                \draw[->, >={Stealth}, color=blue] (0.655, 0.49) to (0.665, 0.475); 
                \node () at (0.35,0.2) {$M$};
            \end{scope}

    \end{tikzpicture}
    \caption{(a): The brick wall circuit ansatz used in this work. The state $\ket{\psi_f}$ is parameterized by $\mathcal{D}$ layers of nearest-neighbor unitary matrices applied to a product state $\ket{\psi_i}$. Each layer consist of two sub-layers acting on even and odd bonds respectively. (b): The necessary unitary condition for the quantum gates $U^\dagger U = \mathbb{I}$. (c): An inversion symmetric layer. The color gradient symbolizes that some matrices within the brick wall are identified by spatial inversion.
    (d): Optimization on the unitary manifold $M$. The gradient $\nabla$ is lifted to the tangent space $T_U M$. The resulting $\nabla^R$ is used to follow the geodesic with a learning rate $\eta$ reaching $U'$ without leaving the unitary manifold. (e): Symmetry constrained enforced by the explicit charge conservation, shown at the example of the parity $\mathcal{P} = \mathcal{P}^\dagger = \prod Z$.}  
    \label{fig:circuit_schemetic}
\end{figure*}
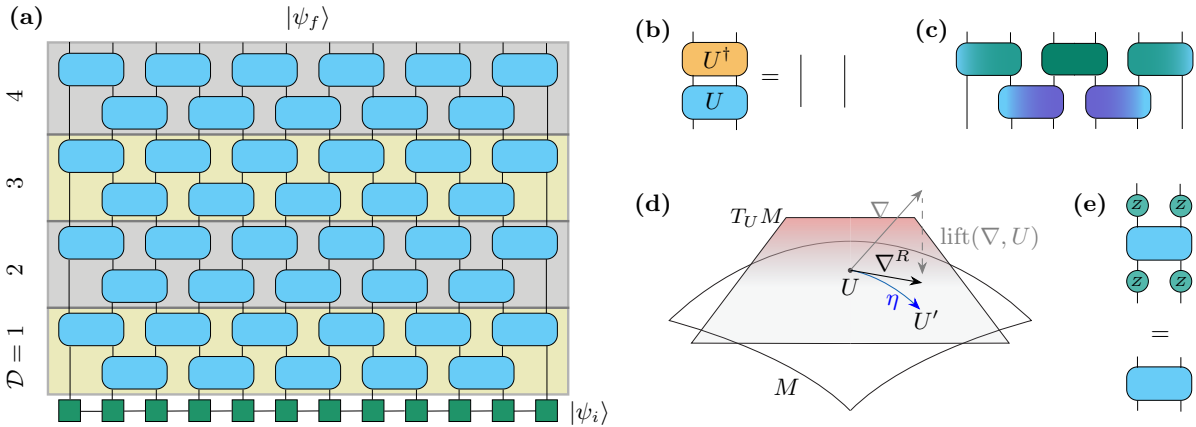
\section{Introduction}
The exponentially-large Hilbert space of a generic quantum many body system poses an insurmountable challenge for the ab-initio investigation of its characteristics using classical computers. Despite the success of tensor network algorithms like time evolution block decimation~(TEBD)~\cite{vidalEfficientSimulationOneDimensional2004a} and density matrix renormalization group~(DMRG)~\cite{whiteDensityMatrixFormulation1992a, schollwoeckDensitymatrixRenormalizationGroup2011a} in simulating low-energy properties of numerous one and two-dimensional lattice  models~\cite{Hastings2007, Vidal2008, Schuch2008, Verstraete2008,Verstraete:2004cf, Schuch2007, Zaletel2020}, analysis of generic highly-entangled states remains out of reach. These states are relevant for characterization of non-equilibrium dynamics as well as low-energy properties of two and three dimensional quantum systems~\cite{jordanClassicalSimulationInfiniteSize2008b,verstraeteRenormalizationAlgorithmsQuantumMany2004,vlaarSimulationThreedimensionalQuantum2021}.

Quantum simulation~\cite{Feynman_1982} promises to be an indispensable tool for the analysis of physical problems that involve the aforementioned highly-entangled states. While paradigmatic quantum algorithms~\cite{Nielsen_Chuang_2000} like phase-estimation~\cite{Kitaev:1995qy} are exponentially-superior compared to their classical counterparts, their implementation in current or near-term quantum simulators to solve a problem of widespread interest remains a daunting challenge. In the near-term, hybrid quantum-classical algorithms, which combine the power of a quantum processor with that of a classical one, are likely to play a central role in solving problems of practical relevance. In fact, the variational quantum eigensolver~\cite{Peruzzo2011, Tilly2022} and its modifications~\cite{Grimsley2019, Tang2021} as well as the quantum approximate optimization algorithm~\cite{Farhi2014, Lloyd2018} have been successful in analyzing quantum chemistry~\cite{Kandala2017} and combinatorial optimization~\cite{Harrigan2021} problems.

The crucial primitives of a hybrid quantum-classical algorithm are a parameterized quantum circuit ansatz and an optimization scheme. The problem of finding the optimal quantum circuit reduces to one in geometric control theory~\cite{Jurdjevic1997, Nielsen2006}. Here, we analyze this problem for the case when the quantum circuit ansatz is constructed using nearest-neighbor operators belonging in SU($d^2$)~[see Fig.~\ref{fig:circuit_schemetic}a)], where~$d$ is the dimension of the local Hilbert space. The models analyzed in this work involve the conventional case of~$d = 2$~(qubits) as well as~$d = 3$~(qutrits). The core of the optimization scheme relies on the flexible automatic differentiation technique~\cite{rumelhartLearningRepresentationsBackpropagating1986,baydinAutomaticDifferentiationMachine}. This enables calculation of gradients of TEBD-based quantum circuit evolutions using matrix product states~(MPS)~\cite{perez-garciaMatrixProductState2007, Schollwock2011}. The optimization is performed by modifying the well-known ADAM optimizer~\cite{kingmaAdamMethodStochastic2014} to work on the smooth SU($d^2$) manifold~\cite{brantnerGeneralizingAdamManifolds2023, becigneulRIEMANNIANADAPTIVEOPTIMIZATION2019}.

While automatic differentiation has been used in the context of quantum circuit optimization~\cite{bergholmPennyLaneAutomaticDifferentiation2022, luchnikovQGOptRiemannianOptimization2021,broughtonTensorFlowQuantumSoftware2021, efthymiouQiboFrameworkQuantum2021} and tensor networks~\cite{liaoDifferentiableProgrammingTensor2019, gengDifferentiableProgrammingIsometric2022, novikovAutomaticDifferentiationRiemannian2021}, attempts to differentiate through tensor network based circuit simulations are sparse. The approach closest to the method presented here is Ref.~\cite{zhangTensorCircuitQuantumSoftware2023}. The main distinction is that our method employs optimization on manifolds and is built on a physically-motivated tensor backend~\cite{juthoJuthoTensorKitJl2024} rather then a machine learning backend~\cite{paszkePyTorchImperativeStyle2019, frostigCompilingMachineLearning, abadiTensorFlowSystemLargescale2016}. This not only allows the explicit conservation of charges that arise in many-body quantum systems~\cite{singhTensorNetworkDecompositions2010, singhTensorNetworkStates2011, weichselbaumNonabelianSymmetriesTensor2012}, but include the automatic differentiation through the (truncated) singular value decomposition (SVD)~\cite{seegerAutoDifferentiatingLinearAlgebra2019}. Derivatives of the SVD allow to optimization approximate circuit evaluations and more complex cost functions like the Jozsa-Uhlmann subspace fidelity~\cite{uhlmannTransitionProbabilityState1976, jozsaFidelityMixedQuantum1994}, see below.

The paper is organized as follows.
Section \ref{sec:QuantumCicuitOptimizer} provides an overview of the algorithm (\ref{sec:optimizeroverview}), followed by a description of the circuit ansatz (\ref{sec:circuitAnsatz}), the optimization on the manifold of unitary matrices (\ref{sec:manifoldOpt}) and the different cost functions used (\ref{sec:costFunctions}). Section \ref{sec:overviewModels} presents three different lattice Hamiltonians: the Ising, the massive Schwinger and the three-state Potts chain. Results for ground and excited states are presented in Secs.~\ref{sec:groundstates} and~\ref{sec:excitedStates} respectively. The details of  automatic differentiation and its implementation for the TEBD-based quantum circuit evolution algorithm is presented in Appendix~\ref{app:AutoDiff}.

\section{The Quantum Circuit Optimizer} \label{sec:QuantumCicuitOptimizer}
\subsection{Overview of the Algorithm} \label{sec:optimizeroverview}

The core idea is to parameterize a quantum circuit as $\mathcal{D}$ layers of nearest-neighbor unitary gates $U_i$. Initially the gates are set as random unitary matrices close to the identity. These gates are applied to an initial quantum state in MPS form $\ket{\psi_i}$ using the TEBD algorithm \cite{vidalEfficientSimulationOneDimensional2004a}.
\begin{align}
    \ket{\psi_f} = \prod_{j=1} U_j \ket{\psi_i} 
\end{align}
The resulting MPS is used to calculate one of three different scalar cost function $\mathcal{L}(\ket{\psi_f})$~(see Sec.~\ref{sec:costFunctions}).

The optimal quantum circuit is obtained by minimizing the relevant cost function using the ADAM optimizer on the manifold of unitary matrices. To that end, we use automatic differentiation~\cite{baydinAutomaticDifferentiationMachine, seegerAutoDifferentiatingLinearAlgebra2019, gengDifferentiableProgrammingIsometric2022, liaoDifferentiableProgrammingTensor2019} which allows efficiently computation of the gradient of the cost function with respect to the unitary matrices. This step is repeated trying to reach the target tolerance. If the cost function saturates without reaching the target tolerance for a given number of layers~$\mathcal{D}$, the circuit ansatz is generalized by increasing $\mathcal{D}$ by 1. The {\it entire}~$\mathcal{D}+1$ layers are subsequently optimized. To improve stability of the layer-growing process, the previously-obtained results, augmented by an initial choice of randomly-initialized unitary operators for the last layer, are used as the initial guess.
The full procedure is shown in Algorithm~\ref{alg:optimizer}.

\begin{algorithm}
    \caption{Optimization Algorithm}
    \label{alg:optimizer}
 \hrulefill\\\vspace{3pt}
    \begin{algorithmic}
        \Require\\
                cost function: $\mathcal{L}(\ket{\psi_f})$\\
                initial state: $\ket{\psi_i} \in \mathrm{MPS}$\\
                tolerance: $\mathrm{tol}$\\
                sub tolerance: $\mathrm{subtol}$\\
                maximum allowed layer depth: $\mathrm{maxdepth}$\\
        \Ensure $\mathcal{L}(\ket{\psi_f}) < \mathrm{tol}$
        \State $\mathcal{D} = 1$ \Comment{initial circuit depth}
        \For{$\mathcal{D} \leq \mathrm{maxdepth}$}
            \State $\ket{\psi_f} = \mathrm{TEBD}(\ket{\psi_i}, \vec{U})$
            \State evaluate $\mathcal{L}\left( \ket{\psi_f} \right)$
            \State evaluate  $\nabla_i = \frac{\partial}{\partial U_i} \mathcal{L}$ \Comment{using AutoDiff}
            \State $\nabla^R_i = \frac{1}{2} \left(U_i^\dagger\nabla_i - \nabla_i^\dagger U_i  \right)$
            \State $\tilde{\nabla}^R = \mathrm{ADAM}(\nabla^R)$ 
            \Comment{manifold ADAM}
            \State $U_i = U_i \exp \left( - \eta \tilde{\nabla}^R_i \right)$
            \If{$\mathcal{L} < \mathrm{tol}$}
                \State break
            \ElsIf{ $|\mathcal{L} - \mathcal{L}_{-1}| < \mathrm{subtol}$}
                \State $\mathcal{D} = \mathcal{D} + 1$ \Comment{increase circuit depth}
                \State reuse old $\vec{U}$ as initial guess
                \State reinitialize $\mathrm{ADAM}$
            \EndIf
        \EndFor
    \end{algorithmic}
 \hrulefill\\\vspace{3pt}
\end{algorithm}
\subsection{The Quantum Circuit Ansatz} 
\label{sec:circuitAnsatz}
We chose as the circuit ansatz a dense brick-wall layout of nearest-neighbor unitary matrices~[Fig.~\ref{fig:circuit_schemetic}(a)], which has been proven successful in the context of quantum circuit optimization~\cite{royUniversalEulerCartanCircuits2024, royEfficientQuantumCircuits2023, bravo-prietoScalingVariationalQuantum2020, jobstFinitedepthScalingInfinite2022a}. It comes with the benefit of only local interactions, which is suitable for many quantum processors. At the same time, it allows efficient classical simulation of the circuit evolution on a classical computer using the TEBD algorithm~\cite{vidalEfficientSimulationOneDimensional2004a}.

Compared to the more common ansätze we have two main distinctions. First,  the gates can be specified to act on qubits but we also have the option to target an arbitrary qudit system. In this work, we present the 3-state Potts model as an example for qutrits. Second, we parameterize the unitary matrices not by a universal gate set or a set of generators of a Lie algebra~\cite{royEfficientQuantumCircuits2023,farhiQuantumApproximateOptimization2014a}, but instead optimize over the full space of nearest-neighbor unitary matrices. That way we avoid the need to guess the specific gates, or circumvent any additional steps in determining the optimal gate choice for an update~\cite{zhuAdaptiveQuantumApproximate2022}.
This is similar to the approach used in Ref.~\cite{royUniversalEulerCartanCircuits2024} where the full space of nearest-neighbor gates is achieved by a parametrization in terms of the Euler-Cartan decomposition~\cite{tucciIntroductionCartanKAK2005, vatanOptimalQuantumCircuits2004, vidalUniversalQuantumCircuit2004}. 
However, in contrast to Ref.~\cite{royUniversalEulerCartanCircuits2024}, we ensure that the operators of the quantum circuit remain on the $\mathrm{SU}(d^2)$ manifold during the optimization using a variation of ADAM optimizer on unitary manifolds, see Sec.~\ref{sec:manifoldOpt} for details.

As shown below, the utilization of symmetries existing for a given physical problem substantially improves the performance of the optimization process. We impose two types of symmetries on the unitary matrices. First, we impose spatial symmetries within the brick wall structure. Here, only inversion symmetry, see Fig. \ref{fig:circuit_schemetic} c), is used in some of the data presented, but translation symmetry is another possible choice. Second, we impose symmetries that originate from operators corresponding to the particle number and parity conservation of the corresponding Hamiltonian. This is achieved by parameterizing the unitary operators by their respective irreducible representation~\cite{singhTensorNetworkDecompositions2010, singhTensorNetworkStates2011, weichselbaumNonabelianSymmetriesTensor2012}, as commonly used by quantum tensor network codes like \textit{TensorKit.jl}~\cite{juthoJuthoTensorKitJl2024}. This led to a  reduction of the number of free parameters leading to a speedup of the optimization process. The circuit depth,~$\mathcal{D}$, is a `hyperparameter' that is determined dynamically within the optimization procedure, see Algorithm \ref{alg:optimizer}.

\subsection{Optimization on the Manifold of Unitary Matrices} \label{sec:manifoldOpt}
 
Using the automatic differentiation method we calculate the derivative of the cost function with respect to all unitary operators $U_i$ of the ansatz.  The so-called reverse mode automatic differentiation allows us to calculate the gradient 
\begin{align}
    \nabla_i = \frac{\partial \mathcal{L}}{\partial U_i}
\end{align}
up to numerical precision. The matrix derivative should be understood as a shorthand notation to a derivative with respect to all the matrix components. The gradient evaluation adds a cost maximally in the order of the circuit evaluation~\cite{baydinAutomaticDifferentiationMachine} and is independent of the number of parameters. This is  different from the approach used in Ref.~\cite{royUniversalEulerCartanCircuits2024} where the explicit derivatives were manually implemented for quantum natural gradient based optimization. 

The matrix derivative is computed by implementing an additional function for each elementary operation used in the program, the so-called pullback or reverse-rule. This pullback function defines how small deviations of the output are propagated to deviation in the input. An automatic differentiation engine like \textit{Zygote.jl} combines the pullbacks using the chain rule of derivatives to calculate the gradient of a complex function. In principle, only pullbacks for fundamental operations need to be defined but higher order pullbacks for linear algebra~\cite{seegerAutoDifferentiatingLinearAlgebra2019}, and tensor routines~\cite{gengDifferentiableProgrammingIsometric2022, liaoDifferentiableProgrammingTensor2019} increase the computational efficiency. For details and optimizations in the case of the TEBD algorithm see appendix \ref{app:AutoDiff}.

Performing an ordinary gradient descent using the so-calculated derivatives would destroy the unitarity condition of the matrices constituting the quantum circuit. To circumvent this problem, we resort to methods developed in the context of differential geometry and the optimization on Riemann manifolds \cite{kianiProjUNNEfficientMethod2022, liEfficientRiemannianOptimization2020}. The group elements of $\mathrm{SU}(d^2)$ are on a Riemannian manifold in the higher dimensional space of the general~$d^2 \times d^2$-matrices. To perform the optimization on this manifold rather than the entire~$d^2 \times d^2$ dimensional space, the Cartesian matrix valued derivatives $\nabla_i$  need to be projected or \textit{lifted} on to the tangent space $T_{U_i} M$ of the manifold $M$
\begin{align}
    \nabla^R_i  = \mathrm{lift}(\nabla_i) = \frac{1}{2} \left(U_i^\dagger\nabla_i - \nabla_i^\dagger U_i  \right).
\end{align}
We call the lifted gradient $\nabla^R_i$ the Riemann gradient.
These Riemann gradients can be used to calculate the updated $U'$ by following the geodesic on the manifold
\begin{align}
    U'_i = U_i \exp \left( - \eta \nabla^R_i \right),
\end{align}
where $\eta$ is the learning rate. See Fig.~\ref{fig:circuit_schemetic}(d) for an illustration.

    To avoid lengthy and problem dependent optimization of the learning rate $\eta$, we resort to the robust and widely-successful ADAM optimizer~\cite{kingmaAdamMethodStochastic2014}. ADAM has two main components. First, it keeps track of the mean (first moment) of previous gradients to modify the current gradient. This is known and analogous to the momentum of a classical mechanics system. Second, it also keeps track of the variance (second moment) of previous gradient components to reduce the step size of highly fluctuating directions.
    Some details of the ADAM optimizer need adjustment to work on Riemann manifolds~\cite{brantnerGeneralizingAdamManifolds2023}. The first and second moments are calculated from the previous Riemann gradients $\nabla_R$ in contrast to their Cartesian counter parts. The scale of complex valued gradients are adjusted for the real and imaginary part individually, so their corresponding variances are stored separately.

Importantly, every operation in the manifold ADAM optimizer is fully compatible with the explicit charge conservation of physics-motivated tensor network libraries. This means the implementation of the optimization on the submanifold of charge conserving unitaries is straightforward.

\subsection{Cost Functions} 
\label{sec:costFunctions}

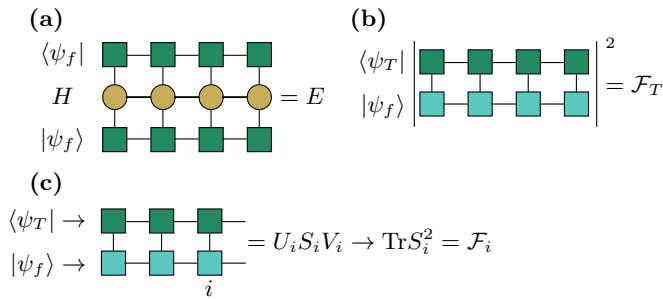
\begin{figure}
    \centering
    \begin{tikzpicture}
        \node[anchor=south west,inner sep=0] (image) at (0,0) {\includegraphics[width=\columnwidth]{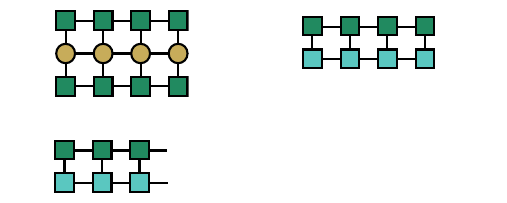}};
        \begin{scope}[x={(image.south east)},y={(image.north west)}, >={Stealth}]
            \node at (0.52, 1.03) {\textbf{(b)}};
            \node at (0.025, 1.03) {\textbf{(a)}};
            \node at (0.025, 0.40) {\textbf{(c)}};
            \node at (0.05, 0.9) {$\bra{\psi_f}$};
            \node at (0.05, 0.74) {$H$};
            \node at (0.05, 0.57) {$\ket{\psi_f}$};
            \node[anchor=west] at (0.37, 0.74) {$=E$};
            \draw (0.87, 0.62) -- (0.87, 0.95);
            \draw (0.59, 0.62) -- (0.59, 0.95);
            \node[anchor=south west] at (0.87, 0.85) {$^2$};
            \node[anchor = east] at (0.59, 0.87) {$\bra{\psi_T}$};
            \node[anchor = east] at (0.59, 0.70) {$\ket{\psi_f}$};
            \node[anchor = west] at (0.87, 0.78) {$=\mathcal{F}_T$};

            \node[anchor=east] at (0.1, 0.26) {$\bra{\psi_T} \rightarrow$};
            \node[anchor=east] at (0.1, 0.09) {$\ket{\psi_f} \rightarrow$};
            \node at (0.275, 0) {$i$};
            \node[anchor = west] at (0.32, 0.18) {$= U_i S_i V_i  \rightarrow \mathrm{Tr} S_i^2 = \mathcal{F}_i$};
        \end{scope}
    \end{tikzpicture}
    \caption{The energy expectation value  (a), the total fidelity (b),  and the subspace fidelity (c) shown in tensor network notation.}
    \label{fig:cost_functions}
\end{figure}

The use of automatic differentiation allows to switch relatively easy between different cost functions, as there is no need to rederive the gradient analytically. 
In this work, we used three cost functions, which are all efficiently calculable if the state is in MPS form.
The most natural choice for the optimization of ground states is the minimization of the energy 
\begin{align}
    E = \braket{\psi_f | H | \psi_f}. \label{eq:costFunctionEnergy}
\end{align}
This is  efficiently implemented by representing the Hamiltonian as a matrix product operator, see Fig.~\ref{fig:cost_functions}(a). The energy is often the desired choice for the cost function since it can be minimized using the variational principle for Hermitian Hamiltonians~\cite{sakuraiModernQuantumMechanics2011}. The Hamiltonian is also relatively easily implemented on real quantum hardware. If the quantum state is already known from classical computations and the goal is to find an efficient quantum circuit for its preparation on quantum hardware, there is more freedom in the choice of the cost function $\mathcal{L}$. In particular, we consider the total fidelity to the target state
\begin{align}
    \mathcal{F}_T = | \braket{\psi_T| \psi_f} |^2 
\end{align}
or its variant the negative log total fidelity  
\begin{align}
    - \log \mathcal{F}_T = -\log | \braket{\psi_T| \psi_f} |^2.\label{eq:totalFid}
\end{align}
The fidelity has been used successfully to prepare states before~\cite{royEfficientQuantumCircuits2023} and its implementation is computationally inexpensive. This is because we consider the case when the state is available as an MPS, see Fig.~\ref{fig:cost_functions}(b). The use of the $- \log \mathcal{F}_T$ is inspired by a similar replacement in the  maximum likelihood optimizations encountered in numerical statistics. The log increases convergence speed especially for big systems. This can be understood analogous to its classical counter part. Two random product states have a overlap that decreases exponentially with system size. This means that for big system the initial overlap and its gradients reach the limit of numerical precision. The logarithm compensates this $\mathcal{F}_T \ll 1$ regime while having no influence for $\mathcal{F}_T \approx 1$. For this reason, every time we describe the results for the total fidelity as a cost function, we performed the minimization on the negative log total fidelity, but omit this detail for brevity. 

Last but not least, we used the the Jozsa-Uhlmann subspace fidelities~\cite{uhlmannTransitionProbabilityState1976, jozsaFidelityMixedQuantum1994}. The latter are defined as
\begin{align}
    \mathcal{F}_i &= \mathcal{F}(\rho^T_i, \rho^C_i )\\
    &= \left( \mathrm{Tr} \sqrt{\sqrt{\rho^T_i} \rho^C_i \sqrt{\rho^T_i}} \right)^2
\end{align}
where the density operator are defined by tracing over the part of the system with lattice site index bigger then $i$:
\begin{align}
    \rho_i = \mathrm{Tr}_{j>i} \ket{\psi} \bra{\psi}.
\end{align}
This cost function is particularly useful for the optimization of states which exhibit certain long-range correlations. In fact, we show that for excited states in physically-relevant models, the subspace fidelity-based cost function outperforms the ordinary total fidelity~(see  Sec.~\ref{sec:excitedStates}). In general, computations of fidelities of density operators are usually expensive, but the special structure of the MPS and the restriction to left-right bipartitions make it computationally feasible, see Fig.~\ref{fig:cost_functions}(c). Compared to full target fidelity only one additional singular value decomposition per subspace fidelity \cite{hauruUhlmannFidelitiesTensor2018} needs to be calculated. This is negligible compared to the cost of the the whole TEBD routine. Importantly, this means the gradient calculation needs to be able to calculate the gradient of this matrix decomposition. This is accomplished by \textit{TensorKit.jl}~\cite{juthoJuthoTensorKitJl2024} together with \textit{Zygote.jl}~\cite{innesDonUnrollAdjoint2018}.
Similar to the log fidelity, we introduce the negative mean log subspace fidelity
\begin{align}
    \bar{\mathcal{F}}_S &= - \frac{1}{L-1} \sum_{i=1}^L \log \mathcal{F}_i(\rho^T_i, \rho^C_i) \label{eq:subFid}
\end{align}
as our last cost function. In the results on subspace fidelity presented below, it is this cost function that is optimized. 
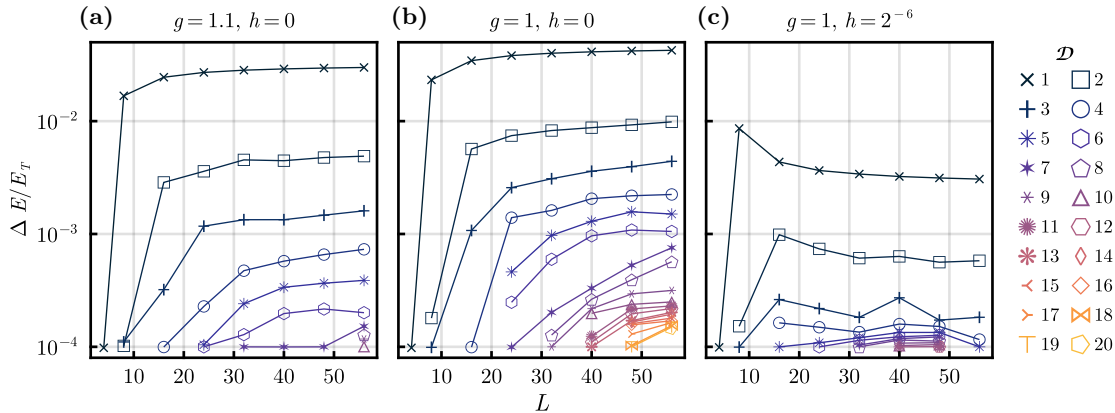
\begin{figure*}
    \centering
    \begin{tikzpicture}
        \node[anchor=south west,inner sep=0] (image) at (0,0) {\includegraphics[width=1.75\columnwidth]{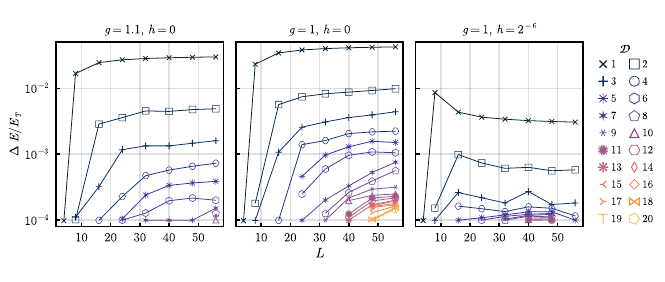}};
        \begin{scope}[x={(image.south east)},y={(image.north west)}, >={Stealth}]
            \node at (0.09, 0.9) {\textbf{(a)}};
            \node at (0.365, 0.9) {\textbf{(b)}};
            \node at (0.632, 0.9) {\textbf{(c)}};
        \end{scope}
    \end{tikzpicture}
    \caption{Relative energy error $\Delta E / E_T$ obtained using the optimized quantum circuit for the Ising model~[Eq.~\eqref{eq:Ising}] ground state. All states are obtained by minimizing the energy expectation value~[Eq.~\eqref{eq:costFunctionEnergy}]. Panels (a) and (c) show the results for two gapped phases with~$g = 1.1, h = 0$ and $g = 1, h = 1/2^6$ respectively. For these cases,~$\mathcal{D} \approx 10$ layers are enough to reach the target accuracy of $\Delta E / E_T = 10^{-4}$. This is nearly independent of the system size $L$. Panel (b) shows the results for the gapless point~$g= 1, h = 0$. In this case, the required number of layers $\mathcal{D}$ to reach a specific energy accuracy increase with system size. Inversion symmetry was used for all three cases while the $\mathbb{Z}_2$ charge was conserved in panels (a, b).}
    \label{fig:Ising_GS_L}
\end{figure*}

\section{Results for eigenstates of lattice quantum field theories.} \label{sec:results}
\subsection{Overview of the Lattice Models} 
\label{sec:overviewModels}
In this section, we apply the introduced algorithm to find shallow quantum circuits that map trivial product states to eigenstates of lattice realizations of 1+1D quantum field theories. The chosen models explore not only those realized using qubits with nearest-neighbor Hamiltonian couplings (like the Ising model), but also all-to-all couplings (like the massive Schwinger model). To demonstrate the power of this approach to optimize circuits based on arbitrary~$SU(d^2)$ operators, we analyze also the three-state Potts model which corresponds to the case~$d = 3$.

The three pertinent lattice Hamiltonians are described below.
\begin{enumerate}[label=\roman*)]
    \item Ising model:
    
        The quantum Ising model with \changed{}{and without} longitudinal field is defined by 
        \begin{align}
            H_\mathrm{Ising} = -  \sum_{i=1}^{L-1} X_i X_{i+1} - \sum_{i=1}^L\left( g Z_i + h X_i\right) \label{eq:Ising}
        \end{align}
        with $X$, $Z$ labeling the Pauli matrices.
        For $h=0$, this exactly solvable model~\cite{kogutIntroductionLatticeGauge1979} has a $\mathbb{Z}_2$ symmetry as its Hamiltonian commutes with the parity operator 
        \begin{align}
            \mathcal{P} = \prod_{i=1}^L Z_i \label{eq:parityOp}.
        \end{align}
        The transition between the ferromagnetic phase~($g<1$) and the paramagnetic phase~($g>1$) occurs at a second order phase quantum critical point. At this point~($g=1$), the low-energy properties of the model are described by a $c=\frac{1}{2}$  conformal field theory (CFT)~\cite{difrancescoConformalFieldTheory1997}. For $h \neq 0, g\leq 1$, the model hosts exotic bound-states of domain-walls. These Ising mesons have an rich mass-spectrum~\cite{McCoy1978, Zamolodchikov1989, Rutkevich2005, Fonseca2006} and have been recently observed on a noisy quantum simulator~\cite{Lamb2023}.
    \item Massive Schwinger model:
    
        The massive Schwinger model~\cite{Coleman:1975pw, Banks1976} can be mapped to a long range spin-$\frac{1}{2}$ chain using a staggered fermion mapping and the Jordan-Wigner transformation \cite{Banuls:2013jaa, farrellScalableCircuitsPreparing2024}
        \begin{align}
            H_\mathrm{MS} = &m\sum^{L}_{i=1} P_i   \label{eq:masSchwing}\\ &+ \frac{1}{2} \sum_{i=1}^{L-1}\left( X_i X_{i+1} + Y_i Y_{i+1} \right)\notag \\ &+\frac{g^2}{2} \sum^{L-1}_{i=1}\left(\sum_{k=1}^i  (-1)^{k} P_k\right)^2 \notag
        \end{align}
        with 
        \begin{align}
            P_k = \frac{1}{2}\left[ \mathbb{I}+ (-1)^k Z_k \right]
        \end{align}
        It commutes with the total spin-z component
        \begin{align}
            Z^\mathrm{tot} = \sum_i^L Z_i
        \end{align}
        giving rise to a $\mathrm{U}(1)$ symmetry, with the ground state lying in the $Z^\mathrm{tot} = 0$ sector.
        This model has been studied numerically \cite{banulsMassSpectrumSchwinger2013} as well as in the context of quantum circuit optimization \cite{farrellScalableCircuitsPreparing2024, royUniversalEulerCartanCircuits2024}.
    \item 3-State Potts Model:
    
        As the last model we consider a qutrit system, the 3-state Potts model with longitudinal field~\cite{rutkevichTwokinkBoundStates2010, rutkevichBaryonMassesThreestate2015, lencsesConfinementQstatePotts2015}
        \begin{align}
            H_\mathrm{Potts} = &- \sum_{i=1}^{L-1} \left( \sigma_i \sigma_{i+1}^\dagger + \sigma^\dagger_i\sigma_{i+1} \right) \label{eq:3StatePotts}\\
            &-g\sum_{i=1}^{L} \left(\tau_i+\tau_{i}^\dagger\right) \notag \\
            &-h \sum_{i=1}^{L} \left(\sigma_i+\sigma_{i}^\dagger\right) \notag
        \end{align}
        with
        \begin{align}
            &\sigma = \begin{pmatrix}
                        0 & 1 & 0\\
                        0 & 0 & 1\\
                        1 & 0 & 0\\
                        \end{pmatrix}
            &\tau = \begin{pmatrix}
                        1 & 0 & 0\\
                        0 & \omega & 0\\
                        0 & 0 & \omega^2\\
                        \end{pmatrix}
        \end{align}
        and the commutation relations
        \begin{align}
            &\sigma\tau = \omega \tau\sigma &\omega  = \exp\left(\frac{2\pi}{3} i \right)
        \end{align}
        The model is a three-state generalization of the Ising case, with a ferromagnetically interacting term, a transverse field with strength $g$ and a longitudinal field with strength $h$. Similar to the Ising model it has a second order phase transition at $h=0$, $g=1$. At this point the low energy physics are described by a $c=\frac{4}{5}$ CFT~\cite{mongParafermionicConformalField2014a}.
        While this model has a global $\mathbb{Z}_3$ for $h=0$ and a $\mathbb{Z}_2$ symmetry for $h\neq 0$, we did not conserve the corresponding charges in the following results.
\end{enumerate}
For these three models, eigenstates in MPS form and eigenenergies are found on a classical computer using DMRG~\cite{whiteDensityMatrixFormulation1992a} and  quasiparticle ansatz method using the \textit{MPSKit.jl}~\cite{vandammeMPSKit2024, haegemanVariationalMatrixProduct2012} library.

\subsection{Ground state optimization}\label{sec:groundstates}

First, we focus on the optimization of circuits preparing ground states.
The latter can be found by the variationally minimizing the cost function corresponding to the expectation value of the Hamiltonian~[Eq.~\eqref{eq:costFunctionEnergy}]. To measure the accuracy, we calculate the relative error 
\begin{align}
    \frac{\Delta E}{E_T} = \frac{\left| \bra{\psi_C} H \ket{\psi_C} - E_T \right|}{E_T}
\end{align}
of the optimized energy to the target energy $E_T$ obtained by DMRG. We also employ this measure as our convergence criterion. In this case, we consider $\Delta E /E_T < 10^{-4}$ as the target tolerance.

\begin{figure}
    \centering
    \begin{tikzpicture}
        \node[anchor=south west,inner sep=0] (image) at (0,0) {\includegraphics[width=\columnwidth]{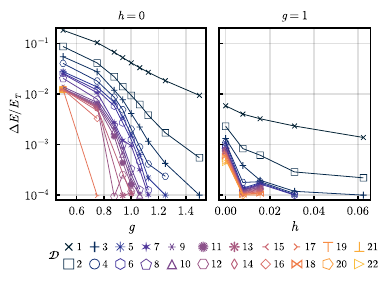}};
        \begin{scope}[x={(image.south east)},y={(image.north west)}, >={Stealth}]
            \node at (0.165, 0.94) {\textbf{(a)}};
            \node at (0.6, 0.94) {\textbf{(b)}};
        \end{scope}
    \end{tikzpicture}
    \caption{Relative energy error $\Delta E /E_T$ for the Ising [Eq.~\eqref{eq:Ising}] ground state of size $L=40$. The circuits where optimized for different parameter values $g$ and $h$ using the energy as the cost function [Eq.~\eqref{eq:costFunctionEnergy}]. (a) While going form the ferromagnetic phase ($g<1$) to the paramagnetic phase ($g>1$) the required layers $\mathcal{D}$ to reach the target energy error decreases. The surprisingly deep necessary circuit depth of the gapped ferromagnetic phase can be understood by the fact that the conserved parity forces the system into a macroscopic superposition. (b) For bigger $h$, only few layers are needed to reach accurate states. Close to and at the critical point additional layers $\mathcal{D}$ give diminishing returns. This is in contrast to (a). This can be understood by the lack of $\mathbb{Z}_2$ charge conservation compared to (a).}
    \label{fig:Ising_GS_lambda}
\end{figure}
Fig.~\ref{fig:Ising_GS_L} shows the relative error in the ground state energy for the Ising model~[Eq.~\eqref{eq:Ising}] for different system sizes $L$ and circuit depths $\mathcal{D}$. Panels (a), (b) and (c) show the results for the paramagnetic gapped phase, the critical point and perturbed from the critical point via the  longitudinal field respectively. In the gapped phases [panels (a,c)], the number of layers required for the same accuracy saturates. For example, for $L\in[36,48]$ of~(c), the number of layers to reach the target accuracy is independent of the system size. At the critical point in b) on the other hand the circuit depth has to increase with the system size to get a comparably accurate state. This is similar to the required increase of the bond dimension which govern the complexity of MPS-based DMRG computations~\cite{pollmannTheoryFiniteEntanglementScaling2009a}, and in agreement with earlier works \cite{jobstFinitedepthScalingInfinite2022a, royEfficientQuantumCircuits2023, royUniversalEulerCartanCircuits2024}.

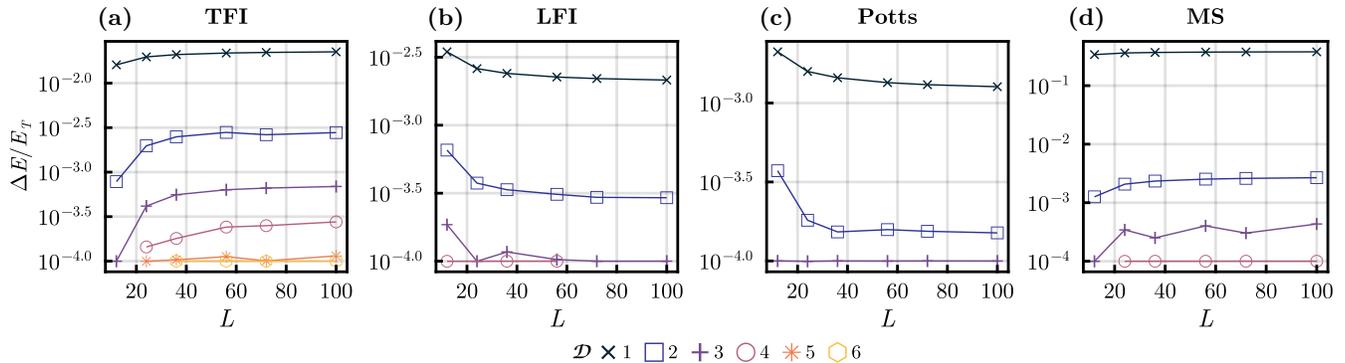
\begin{figure*}
    \centering
    \begin{tikzpicture}
        \node[anchor=south west,inner sep=0] (image) at (0,0) {\includegraphics[width=\textwidth]{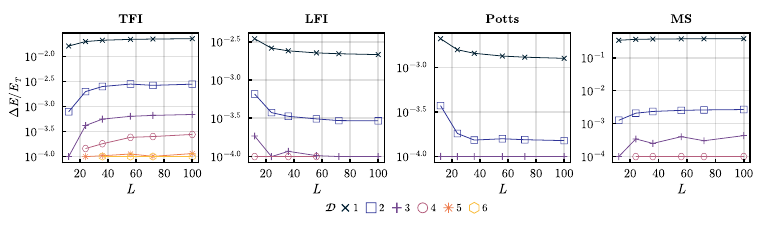}};
        \begin{scope}[x={(image.south east)},y={(image.north west)}, >={Stealth}]
            \node at (0.09, 0.91) {\textbf{(a)}};
            \node at (0.335, 0.91) {\textbf{(b)}};
            \node at (0.58, 0.91) {\textbf{(c)}};
            \node at (0.81, 0.91) {\textbf{(d)}};
        \end{scope}
    \end{tikzpicture}
    \caption{Relative energy error $\Delta E / E_T$ for different system sizes up to $L=100$. For all four models the parameters where chosen such that they require a similar circuit depth $\mathcal{D}$ to reach the target accuracy of $10^{-4}$ by minimizing the circuit energy [Eq.~\eqref{eq:costFunctionEnergy}]. The transverse field Ising model [Eq.~\eqref{eq:Ising}] with $g=1.2, h = 0$ and Ising model with transverse and longitudinal field $g=1, h = 0.03$ is shown in (a) and (b). The 3-state Potts model [Eq.~\eqref{eq:3StatePotts}] with $g=1, h=0.03$ and massive Schwinger model [Eq.~\eqref{eq:masSchwing}] with $g = 0.3, m=0.5$ is shown in (c) and (d). The total parity was conserved in the TFI and the total spin z-component for the MS case.}
    \label{fig:ground_state_all_systems_big}
\end{figure*}
Even though the aforementioned growth in circuit depth at the critical point such as the one for~$g\rightarrow 1^+, h = 0$ is similar to that of the bond-dimension in an MPS-based DMRG computations, there is an important distinction. This is visible in the results obtained for the ferromagnetic regime~$g<1, h = 0$. In this phase, even though the spectrum is gapped, the ground state is two-fold degenerate. In the extreme case of~$g = h = 0$, the ground state is the `cat state': $\ket{\psi_c}= \frac{1}{\sqrt{2}}\left(\ket{\rightarrow \cdots \rightarrow} + \ket{\leftarrow \cdots \leftarrow}\right)$. In contrast to the bond-dimension of the MPS required to represent this state, the circuit-depth scales with the system-size. This is required to form the nonlocal superposition from local unitary operations. Note that this effect is present only when the optimization is performed while conserving the~$\mathbb{Z}_2$ charge~[Eq.~\eqref{eq:parityOp}]. Of course, if the target state is not imposed to be an eigenstate of the operator~$\mathcal{P}$, the circuit depth is dramatically reduced since no nonlocal superposition needs to form at the end of the quantum circuit evolution. 

The conservation of the above charge during the optimization also improved the efficiency of the proposed scheme. This is seen for the~$g = 1, h = 0$ case in panels (a, b) of Fig.~\ref{fig:Ising_GS_lambda}. In contrast to panel (b), panel (a) presents the results for the optimization performed while conserving the said ~$\mathbb{Z}_2$ charge. The target accuracy was reached in panel (a) after $\mathcal{D} = 14$ layers, while the same is not true even after~$\mathcal{D}=22$ layers in panel (b). We attribute this to the more efficient optimization performed in panel (a) compared to panel (b) due to the charge conservation. This is reminiscent of the benefits of charge-conservation in standard tensor-network based algorithms like DMRG. 

Thus, the proposed optimization scheme finds accurate shallow representation of a 1D gapped ground states for large systems as long as the spectral gap is large enough and no macroscopic superposition needs to form. This is supported also by the results presented in Fig. \ref{fig:ground_state_all_systems_big}. It shows the relative energy error for the Ising model with transverse field in~(a) with transverse and longitudinal field in~(b), the 3-state Potts model in~(c) and the massive Schwinger model in~(d). All four cases are in the gapped phase with small correlation lengths and therefore reach the target energy accuracy of $10^{-4}$ within a few ($\mathcal{D} \leq 6$) layers. This is true even for system sizes up to $L=100$. The ability to reliably probe such large system-sizes emphasizes the efficacy of the proposed scheme. We do not use any  extrapolation techniques~\cite{farrellScalableCircuitsPreparing2024} since the latter are unlikely to be well-controlled and viable for generic models. 

\subsection{Excited States}
\label{sec:excitedStates}
Next, we present results for excited states of the aforementioned models. We optimized circuits for the ground state as well as the first ten~(indexed by~$n$) eigenstates of the models introduced above within their respective groundstate charge sector. In contrast to ground states, excited states cannot be found by simply minimizing the energy. Instead, a modified variational principle needs to be used by projecting out or penalizing the overlap with previously-obtained eigenstates. This method has been used in~\cite{royUniversalEulerCartanCircuits2024}. The optimizer presented here only showed inconsistent success with this approach. This indicates that this kind of cost function likely requires at least a quasi-second order optimizer like the quantum natural gradient~\cite{stokesQuantumNaturalGradient2020,royEfficientQuantumCircuits2023, royUniversalEulerCartanCircuits2024}, instead of the first order ADAM optimizer presented here.

\begin{figure}
    \centering
    \begin{tikzpicture}
        \node[anchor=south west,inner sep=0] (image) at (0,0) {\includegraphics[width=\columnwidth]{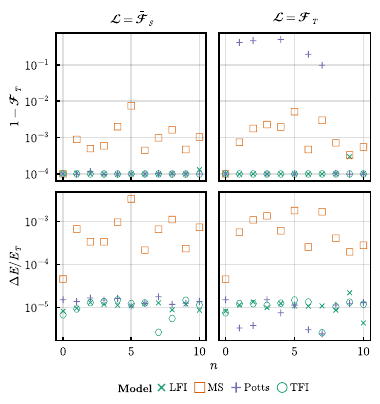}};
        \begin{scope}[x={(image.south east)},y={(image.north west)}, >={Stealth}]
            \node at (0.505, 0.89) {\textbf{(a)}};
            \node at (0.935, 0.89) {\textbf{(b)}};
            \node at (0.505, 0.5) {\textbf{(c)}};
            \node at (0.935, 0.5) {\textbf{(d)}};
        \end{scope}
    \end{tikzpicture}
    \caption{Total infidelity $1-\mathcal{F}_T$ (top) and relative energy error $\Delta E /E_T$ (bottom) for the ground state $n=0$ and the first $n=10$ excited states. The left and right columns show the results when the cost functions were chosen to be the subspace fidelity $\bar{\mathcal{F}}_S$ and the total fidelity $\mathcal{F}_T$ respectively. This comparison is shown for four different cases: the Ising model with transverse field only (TFI) $g=1.2, h = 0$, Ising model with longitudinal and transverse field (LFI) $g=1, h=0.03$, the 3-state Potts $g=1, h=0.5$, and the massive Schwinger (MS) model $g=0.3, m = 0.5$. Both cost functions give excellent agreement for the relative energy error, but only the subspace fidelity gives accurate results for the total infidelity in all cases. For better comparison the total infidelity of $10^{-4}$ was chosen as the convergence criterion independent of the optimized cost function.  All cases have a system size of $L = 24$ and all but the massive Schwinger model reached or got close to the target fidelity within $\mathcal{D} = 16$ layers.} 
    \label{fig:excited_states_cost_func_comp}
\end{figure}
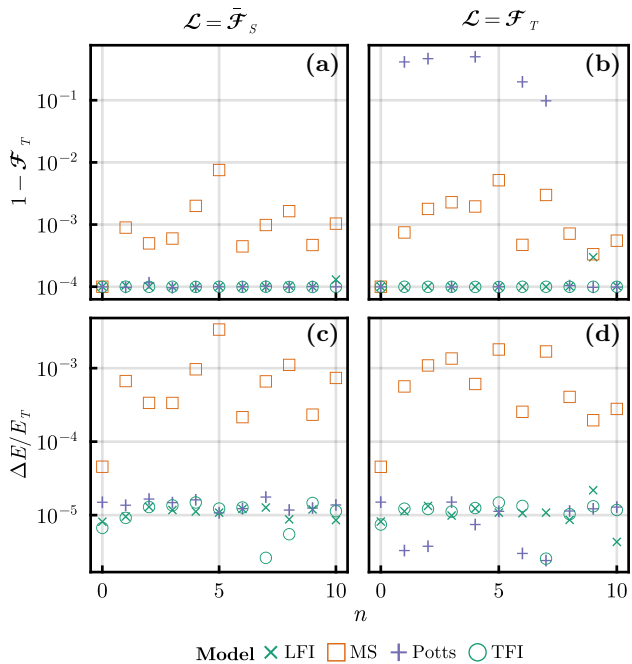

As an alternative, we use the fact that we can calculate the states classically using tensor network methods and target the corresponding state. The natural choice is to use the total fidelity [Eq.~\eqref{eq:totalFid}] as a cost function, but we also propose the mean subspace fidelity [Eq.~\eqref{eq:subFid}].  Fig.~\ref{fig:excited_states_cost_func_comp} shows the infidelity as well as the energy accuracy for the different models and eigenstates for both cost functions. Both cost functions converge to accurate relative energy errors (bottom) within the maximum number of layers $\mathcal{D} = 16$ and in all models. With $\Delta E /E_T \approx 10^{-3}$ the most inaccurate results are  observed for the long range massive Schwinger model.

A closer look to the total infidelity (top) reveals that some states are as badly converged as $0.5$ and far of the total target infidelity of $10^{-4}$. This convergence issue is only observed when using $\mathcal{F}_T$ as the cost function $\mathcal{L}$, rather $\bar{\mathcal{F}}_S$.
Nonetheless, these very `inaccurate' states give remarkably well-converged energies. The cause of the discrepancy in fidelity can be attributed to the formation of nonlocal superpositions of states which differ by a flipped spin at the two open boundaries. For the finite size open systems analyzed in this work, the symmetric superposition has the lower energy. However, such a nonlocal superposition is hard to realize with the local optimization updates. The energies, however, remain close to the target energy. This is because the difference between the energies for the symmetric and anti-symmetric superpositions are exponentially small in the system-size. 

When optimizing using the total target state fidelity cost function, the gradient descent  greedily optimizes only one part of this superposition. After this false minima is reached only local nearest-neighbor updates are performed. The chance that these create a overlap with the other half of the super position and without negatively impacting the already established half, gets lower and lower the further the edge states are separated. The algorithm has a hard time establishing a continuous gradient for this kind of entangled state. On the other hand, the subspace fidelity cost function is already sensitive to the existence of a superposition state by comparing the reduce density operators of the first site. This enables following that gradient and match the second site density operator for sites 1 and 2. In such a way, the subspace fidelity allows to continuously grow long-range entanglement using gradient descent. 

We emphasize that accurately capturing subtle boundary effects are crucial for a wide family of models, notable examples include the Kitaev wire~\cite{Kitaev2001} and the AKLT chain~\cite{affleckRigorousResultsValencebond1987, affleckValenceBondGround1988}. The proposed subspace fidelity-based cost function provides a robust way for realizing these states on a quantum simulator.

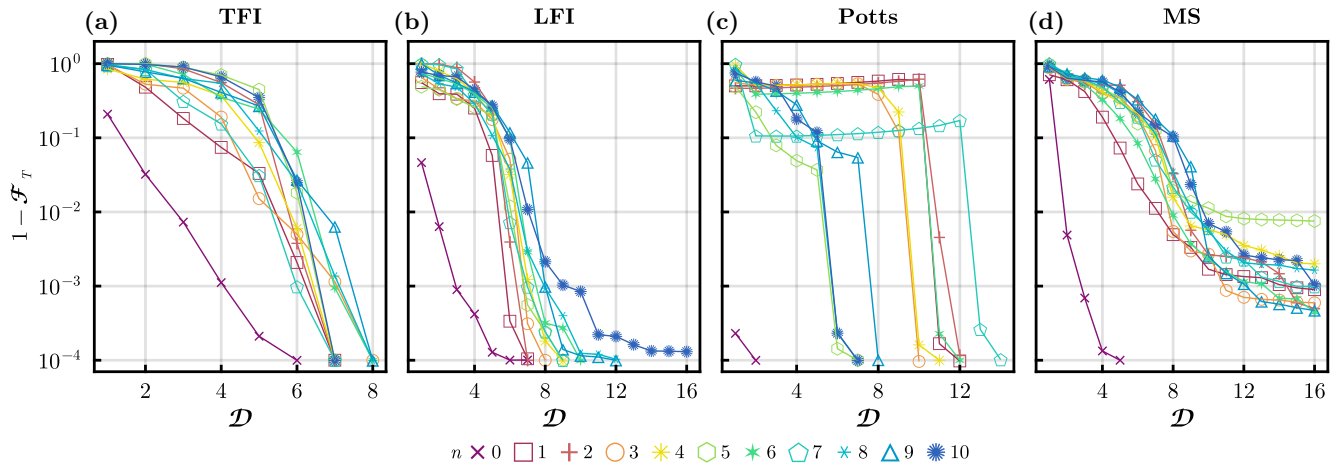
\begin{figure*}
    \centering
    \begin{tikzpicture}
        \node[anchor=south west,inner sep=0] (image) at (0,0) {\includegraphics[width=\textwidth]{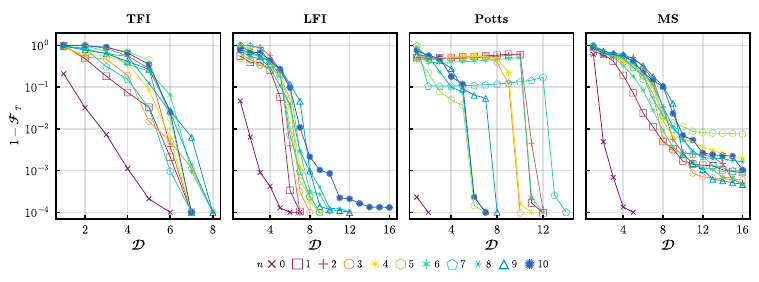}};
        \begin{scope}[x={(image.south east)},y={(image.north west)}, >={Stealth}]
            \node at (0.08, 0.92) {\textbf{(a)}};
            \node at (0.31, 0.92) {\textbf{(b)}};
            \node at (0.54, 0.92) {\textbf{(c)}};
            \node at (0.78, 0.92) {\textbf{(d)}};
        \end{scope}
    \end{tikzpicture}
    \caption{The circuit depth $\mathcal{D}$ dependence for the ground state and the first 10 excited states of the four different cases within their respective ground state charge sector. All systems of size $L=24$ where optimized using $\bar{\mathcal{F}}_S$ as a cost function.
    The model parameter are the same as described in Fig.~\ref{fig:excited_states_cost_func_comp}.
    The ground states (magenta cross) has the most rapid convergence with respect to the circuit depth. While some excited states have a slower but similar behavior, other states show a non monotonous convergence. This is due to the fact that the optimization of the subspace fidelity allows to build up the long range entanglement even for circuits not yet deep enough. This comes at the cost of a reduced total infidelity. Only when the circuit is deep enough to represent the long range entanglement, the total infidelity drops suddenly.
    }
    \label{fig:rest_excited}
\end{figure*}

Fig. \ref{fig:rest_excited} shows the circuit depth dependence for the different eigenstates and models for the subspace fidelity cost function. As expected from the small correlation length, the convergence of the ground state is quite rapid in all models. The excited states require significantly more layers and do not show a clear order or hierarchy with $n$ but often a similar convergence behavior as the ground states. 
A few cases show a plateau like trend with a small incline, until a specific circuit depth is reached and then suddenly drop close to the target fidelity. This includes all cases in which the total fidelity cost function could not reach the target state. Optimizing the subspace fidelity does increase the long range entanglement at the cost of the total fidelity even if the circuit depth cannot reach the target state yet. As the circuit depth reaches $\mathcal{D} \approx L/2$, the circuit allows entanglement between both edges and the total fidelity drops.

\section{Conclusion}
To summarize, in this work, an efficient quantum circuit optimization scheme is presented combining methods of machine learning and tensor networks. A variation of the ADAM optimizer for manifolds of charge-conserving unitary operators is utilized together with reverse mode automatic differentiation for a TEBD-based quantum circuit evolution. The efficacy of this scheme is demonstrated by computing  accurate shallow quantum circuits for the lowest eleven eigenstates of many-body spin-chain Hamiltonians with both local and nonlocal interactions. 

The proposed scheme is crucial for hybrid variational quantum-classical algorithms that rely on an efficient preparation of a given state on a quantum hardware. The so-prepared state can either serve as an initial guess for other quantum algorithms or for the investigation of non-equilibrium dynamics. The latter can probe characteristics of quantum systems that lie beyond the reach of current classical computers. 

The described circuit ansatz can be generalized to improve its performance. A natural extension is to consider qudits on graphs with more general connectivity than nearest-neighbor, potentially tailoring to specific quantum hardware platforms. The proposed optimization scheme is also amenable to GPU and tensor acceleration. This would not only speed up the evaluation times further \cite{GitHubNVIDIACudaquantum, unfriedFastTimeEvolutionMatrixProduct2023a}, but allow a fair comparison with other more traditional machine learning architectures. Finally, inclusion of higher-order optimization routines like the quantum natural gradient~\cite{stokesQuantumNaturalGradient2020, royEfficientQuantumCircuits2023, royUniversalEulerCartanCircuits2024} could further enhance the performance of the proposed scheme. 
\section*{Acknowledgments}
This work was supported by the U.S. Department of Energy, Office of Basic Energy Sciences, under Contract No. DE-SC0012704.
\FloatBarrier

\appendix
\section{Automatic Differentiation}\label{app:AutoDiff}
\subsection{Reverse Mode Automatic Differentiation}
Gradient based optimization algorithms require access to the gradient of the cost function $\mathcal{L}(\theta)$ regarding its input parameters $ \theta$. The traditional way to calculate these derivatives is either analytic or using finite differences. Analytic calculations for Jacobians of more complex functions are laborious, and require a "nice" mathematical expression of the cost function. If the function involves on the output of a complex computer algorithms a analytic expression is often out of reach. Finite differences on the other hand can be applied to arbitrary computer routines, but suffer from precision errors associated to the finite size step length.
Another big issue for finite differences arises if the output dimension is much smaller than the input dimension as it is the case for most minimization problems. Finite differences can only map variations in the input to variations in the output, one perturbs the parameter and see how the cost function changes. To find the gradient, the perturbation of steepest descent, one needs to perturb each input parameter independently. Automatic Differentiation (AD)~\cite{baydinAutomaticDifferentiationMachine} tackles these issues by providing algorithms that calculate derivatives of complex computer programs up to numerical precision nearly automatically.

In the following we will focus on a specific flavor of AD the so called reverse mode AD which is generalization of the back-propagation algorithm known in the context of machine learning. Compared to its counterpart the forward mode differentiation, it is computationally more efficient when the output dimension is smaller then the input dimension. This is satisfied by all the cases we study in this work as we only minimize scalar cost functions. We will assume for simplicity that the differentiable program acts as a real scalar function.
\begin{align}
    f: &\mathbb{R}^n \longrightarrow \mathbb{R}: \;x \xrightarrow{f(x)} y
\end{align}
while we will be working with derivatives of complex numbers $\mathbb{C}^k$ they should be interpreted as a convenient tool to represent multi-variant function of $\mathbb{R}^{2k}$ rather then in the context of complex analysis.
The final goal is to calculate the gradient of this function at an arbitrary point $\tilde{x}$ with respect to all its input parameters:
\begin{align}
    \nabla y|_{\tilde{x}} = \begin{bmatrix}
        \frac{\partial y}{\partial x_1}\Big|_{\tilde{x}} \\
        \frac{\partial y}{\partial x_2}\Big|_{\tilde{x}} \\
        \vdots\\
        \frac{\partial y}{\partial x_{n-1}}\Big|_{\tilde{x}}\\
        \frac{\partial y}{\partial x_{n}}\Big|_{\tilde{x}}
    \end{bmatrix}
    = \begin{bmatrix}
        \bar{x}_1 \\
        \bar{x}_2 \\
        \vdots\\
        \bar{x}_{n-1}\\
        \bar{x}_{n}
    \end{bmatrix}
\end{align}
Here we choose the notation $\bar{v} = \frac{\partial y}{\partial v} \Big|_{\tilde{x}}$ which is often called adjoint of the variable $v$. 
The key observation of the reverse mode differentiation is that the adjoint of the input variables can be computed using the adjoints of the intermediate variables:
\begin{align}
    x &\xrightarrow{g_1(x)} v^1 \xrightarrow{g_2(v^1)} \dots v^{L-1} \xrightarrow{g_L(v^{L-1})} y\\
    \bar{x} &\xleftarrow{g_1^*(\bar{v}^1|x)} \bar{v}^1 \xleftarrow{g_2^*(\bar{v}^{2}|v^1)} \dots \bar{v}^{L-1} \xleftarrow{g^*_L(\bar{y}|v^{L-1})} \bar{y}
\end{align}
where the mappings $g^*_i(\bar{v^i}|v^{i-1})$ are the so-called \textit{pullback} functions or reverse differentiation rules of the corresponding elementary function $g_i(v^{i-1})$.
By using the Leibniz chain rule on the intermediate variables 
\begin{align}
    \bar{y} &= \frac{\partial y}{\partial y}\\
    \bar{v}^{L-1} &= \frac{\partial y}{\partial v^{L-1}} =  \frac{\partial y}{\partial y} \frac{\partial y}{\partial v^{L-1}}\\
    &= \bar{y} \frac{\partial y}{\partial v^{L-1}}\\
    \bar{v}^{L-2} &= \frac{\partial y}{\partial v^{L-2}} =  \frac{\partial y}{\partial y} \frac{\partial y}{\partial v^{L-1}}\frac{\partial v^{L-1}}{\partial v^{L-2}}\\
    &= \bar{v}^{L-1} \frac{\partial v^{L-1}}{\partial v^{L-2}}\\
    &\vdots \notag\\
    \bar{x}&=\bar{v}^1\frac{\partial v^1}{\partial x} 
\end{align}
it is easy to identify pullback functions
\begin{align}
    g^*_n(\bar{v}^{n}|v^{n-1}) &=  \bar{v}^{n-1}\\
    &= \bar{v}^{n}_i\frac{\partial v_i^{n}}{\partial v_j^{n-1}}\bigg|_{v^{n-1}}\\
    &= \left(\mathbf{J}_{g_n}\bigg|_{v^{n-1}}\right)^T \bar{v}^{n}
\end{align}
as the product of the transpose Jacobian (of the function $g_n$ evaluated at $v^{n-1}$) with the vector $\bar{v}^n$ .
Importantly the pullback function only needs to provide an implicit definition of the product, at no point in time is the construction of the full Jacobian necessary.
But the calculation of $\bar{v}^{n-1}$ from $\bar{v}^{n}$ requires a record of which pullback $g_n^*$ connects the two, the intermediate values $v^{n-1}$ to calculate the Jacobian as well as $\bar{v}^{n}$. This means it is necessary to traverse the function twice one in a forward pass, while storing the necessary intermediate results as well as keeping track of the correct variable dependencies, and one backwards pass which uses the correct pullback function to propagate the adjoint back to the input variable. All the information can be stored in a computational graph, this storage of the intermediate results is the main computational limitation for the differentiation of deep routines.

The remaining piece is the definition of the associated pullback function, while it is in principle possible to reduce every operation to it's fundamental operations like addition, multiplication, transcendental functions etc. it can be beneficial to define analytically derived custom higher level pullbacks. A common example are linear algebra routines like matrix multiplications and factorization, this reduces the depth of the computational graph significantly. 

In the following we want to elaborate the crucial simplifications of the computational graph in the context of the TEBD algorithm.

\subsection{Reverse Rule for the TEBD}
In this work we represent the many body wave function in terms of Matrix Product States (MPS). 
Explicitly we choose two sets of tensors, the rank-3 site-tensors 
\begin{align}
    \left\{ A^{(i)}_{\alpha_{i}, s_{i}, \alpha_{i+1}}  \text{ with } 0< i \leq L  \right\}.
\end{align}
and the square rank-2 bond-tensors
\begin{align}
    \left\{ C^{(i)}_{\alpha_{i+1}, \tilde{\alpha}_{i+1}}  \text{ with } 0< i \leq L  \right\}
\end{align}
The indices $s_i$ represent the local Hilbert spaces, $\alpha_i$ label the virtual spaces and are chosen large enough to capture the necessary entanglement of the state faithfully. The only exceptions are $\alpha_1$ and $\alpha_{L+1}$ which are of dimension $1$ as we are restricting ourselves to finite MPS.
The tensors are chosen in the so-called left canonical form such that
\begin{align}
    A^{(n)}_{\alpha_{n}, s_n, \alpha_{n+1}} (A^\dagger)^{(n)}_{\alpha_{n}, s_n, \tilde{\alpha}_{n+1}} = \delta_{\alpha_{n+1}, \tilde{\alpha}_{n+1}}.
\end{align}
While any state can be represented with the $A^{(n)}$s and a single $C^{(L)} = \braket{\psi|\psi}$ via
\begin{align}
    \ket{\psi} = &A^{(1)}_{\alpha_{1}, s_1, \alpha_{2}} A^{(2)}_{\alpha_{2}, s_2, \alpha_{3}} A^{(3)}_{\alpha_{3}, s_3, \alpha_{4}}\cdots \notag \\ &\cdots A^{(L-1)}_{\alpha_{L-2}, s_{L-1},\alpha_{L-1}} A^{(L)}_{\alpha_{L-1}, s_L, \tilde{\alpha}_{L}} C^{(L)}_{\tilde{\alpha}_{L}, \alpha_{L}},
\end{align}
it is convenient to save one $C$-tensor per virtual bond as they allow to switch between right and left canonical form  \cite{schollwoeckDensitymatrixRenormalizationGroup2011a} using
\begin{align}
     B^{(n)}_{\alpha_{n}, s_n, \alpha_{n+1}} &= \left(C^{(n-1)}\right)^{-1}_{\alpha_{n}, \tilde{\alpha}_{n}} A^{(n)}_{\tilde{\alpha}_{n}, s_n, \tilde{\alpha}_{n+1}} C^{(n)}_{ \tilde{\alpha}_{n+1},\alpha_{n+1}} \label{eq:AtoB}\\
     \delta_{\alpha_{n}, \tilde{\alpha}_{n}} &= B^{(n)}_{\alpha_{n}, s_n, \alpha_{n+1}} (B^\dagger)^{(n)}_{\tilde{\alpha}_{n}, s_n, \alpha_{n+1}}.
\end{align}
Any MPS can be brought into this form while also calculate the corresponding $C^{(n)}$s by performing sequential LQ decomposition's from right to the left followed by QR decomposition's from the left to the right.

We use reverse mode automatic differentiation, to calculate gradient of cost functions[See Sec. \ref{sec:costFunctions}] in the form $\mathcal{L}(\psi_f)$
where 
\begin{align}
    \ket{\psi_f} = U\ket{\psi_i}
\end{align}

The calculations of (partial) overlaps and expectation values with matrix product states only consist of tensor contractions where the pullback is well known~\cite{liaoDifferentiableProgrammingTensor2019, gengDifferentiableProgrammingIsometric2022} and recently implemented in some tensor network libraries like \textit{TensorKit.jl}~\cite{juthoJuthoTensorKitJl2024}. Only the subspace fidelity [Eq. \eqref{eq:subFid}] requires additionally the reverse rule for the singular value decomposition which is understood as well~\cite{seegerAutoDifferentiatingLinearAlgebra2019}. For this reason we focus on the application of the unitary operator $U$ using the TEBD algorithm.

In the following it is easier to ignore that the contraction of the tensors result in one quantum state $\ket{\psi}$ but instead think of an MPS as an 1D array $\Psi$ where the entries are alternating $A$ and $C$-tensors.
\begin{align}
    \Psi = \begin{bmatrix}
        A^{(1)} \\
        C^{(1)} \\
        \vdots\\
        A^{(L)}\\
        C^{(L)}
    \end{bmatrix}
\end{align}
The TEBD algorithm applies a unitary $U$ to a state $\ket{\psi}$ in a controlled approximate manner.
But in this interpretation the TEBD algorithm can be understood as a specific mapping from  a unitary $U$ and one tensor array $\Psi$ to another
\begin{align}
    \Phi &= \mathrm{TEBD}(\Psi, U).
\end{align}
The goal is to find the pullback function which maps adjoints in the output $\bar{\Phi}$ to adjoints of the inputs of the $\mathrm{TEBD}$ function
\begin{align}
     \mathrm{TEBD}^*: \bar{\Phi} \longrightarrow   \left(\bar{\Psi}, \bar{U} \right)
\end{align}
The adjoint $\bar{\Phi}$ should be understood as the derivatives of an arbitrary scalar cost function $\mathcal{L}$ with respect to the real $\mathcal{R}$ and imaginary $\mathcal{I}$ part of all the tensor components of $\Phi$:
\begin{align}
    \bar{\Phi}  =
    \begin{bmatrix}
        \frac{\partial \mathcal{L}}{\partial \mathcal{R}({A}_{i,j,k}^{(1)})} +  \frac{\partial \mathcal{L}}{\partial \mathcal{I}({A}_{i,j,k}^{(1)})}\\
        \frac{\partial \mathcal{L}}{\partial \mathcal{R}(C_{i,j}^{(1)})} + \frac{\partial \mathcal{L}}{\partial \mathcal{I}(C_{i,j}^{(1)})}\\
        \vdots\\
        \frac{\partial \mathcal{L}}{\partial \mathcal{R}(A_{i,j,k}^{(L)})} +  \frac{\partial \mathcal{L}}{\partial  \mathcal{I}(A_{i,j,k}^{(L)})}\\
        \frac{\partial \mathcal{L}}{\partial \mathcal{R}(C_{i,j}^{(L)})} + \frac{\partial \mathcal{L}}{\partial  \mathcal{I}(C_{i,j}^{(L)})}
    \end{bmatrix}
\end{align}
similar for $\bar{\Psi}$ and $\bar{U}$.

As described above this pullback function can be constructed efficiently by decomposing the $\mathrm{TEBD}$ function into its fundamental operations and define the  $\mathrm{TEBD}^*$ as an implicit composition of their respective pullbacks.

The TEBD algorithm requires that the applied unitary $U$ is decomposed into layers of bond alternating nearest-neighbor unitary matrices $U^i_j$ where $i$ labels the layer and $j$ each unitary within a layer. This can be achieved using a Suzuki-Trotter decomposition of $U$, or by directly defining the unitary operator $U$ as a network of suitable chosen local quantum gates. The later is used in this work as it is a natural parametrization for quantum circuits. These layers represent one sub-layer introduced in Fig.\ref{fig:circuit_schemetic}(a).
The full TEBD evolution is therefore a composition of multiple TEBDLayer mappings 
\begin{align}
    \Phi &= \mathrm{TEBD}(\Psi, U) \\
    &= \mathrm{Layer}(\nu_{N-1}, U^{N}_j)\\
    \nu_{N-1} &= \mathrm{Layer}(\nu_{N-2}, U^{N-1}_j)\\
    &\;\;\vdots \notag\\
    \nu_{2} &= \mathrm{Layer}(\nu_{1}, U^{2}_j)\\
    \nu_{1} &= \mathrm{Layer}(\Psi, U^{1}_j)
\end{align}
each layer maps a set of non overlapping two-site unitary gates $U_j$ and an array of tensors to another array of tensors:
\begin{align}
    \nu_{i} &= \mathrm{Layer}(\nu_{i-1}, U^i_j)\\
    \mathrm{Layer}&:
    \left(
        \begin{bmatrix}
            A^{(1)}_{i-1} \\
            C^{(1)}_{i-1} \\
            \vdots\\
            A^{(L)}_{i-1}\\
            C^{(L)}_{i-1}
        \end{bmatrix},
        \begin{bmatrix}
            U_1^i \\
            U_2^i \\
            \vdots\\
        \end{bmatrix}
    \right)\longrightarrow
    \begin{bmatrix}
        {A}^{(1)}_i \\
        {C}^{(1)}_i \\
        \vdots\\
        {A}^{(L)}_i\\
        {C}^{(L)}_i
    \end{bmatrix}
\end{align}
where $U_j^i$ are the gates applied on the respective layer.
The fact that $\mathrm{TEBD}$ is a simple composition of $\mathrm{TEBDLayer}$ function implies that the pullback function $\mathrm{TEBD}^*$ is again a composition of $\mathrm{TEBDLayer}^*$ functions.
While the adjoints of the state $\bar{\nu}_i$ are passed on from pullback to pullback the $\bar{U^i}$ will be obtained layer by layer.

Finally the $\mathrm{TEBDLayer}$ can again be decomposed into individual $\mathrm{TEBDSteps}$ each only acting one two-site unitary $U_{(k,k+1)}$ and the subset of the tensors $[A^{(k)}, C^{(k)}, A^{(k+1)}, C^{(k+1)}]$ in layer $i$:
\begin{align}
    \mathrm{TEBDStep}: 
    \left(
        \begin{bmatrix}
            A^{(k)}_{i-1} \\
            C^{(k)}_{i-1} \\
            A^{(k+1)}_{i-1}\\
            C^{(k+1)}_{i-1}
        \end{bmatrix},U_{(k,k+1)}
    \right)\longrightarrow 
        \begin{bmatrix}
        \tilde{A}^{(k)}_{i} \\
        \tilde{C}^{(k)}_{i} \\
        \tilde{A}^{(k+1)}_{i}\\
        \tilde{C}^{(k+1)}_{i}
    \end{bmatrix}
\end{align}
As a single $\mathrm{TEBDLayer}$ only acts on even or odd bonds respectively, no tensor is updated twice within a single layer.
This not only allows to easily parallelize the computation but also implies that the Jacobian of the $\mathrm{TEBDLayer}$ mapping with respect to the input tensors is block diagonal within each $\mathrm{TEBDStep}$ block. 
In other words the pullback of each $\mathrm{TEBDLayer}^*$ is a concatenation of the smaller pullbacks $\mathrm{TEBDStep}^*$ each acting on four tensors only.
This simplification is overlooked in many automatic differentiation engines, like \textit{Zygote.jl} which is used in this implementation, as they do not track vector elements individually in the computational graph. A manual implementation of this specific reverse rule reduces the depth of the computational graph by a factor of half the physical system size times the circuit depth and therefore the memory footprint of the AD routine dramatically.
Further dissecting the individual steps of the $\mathrm{TEBDStep}^*$ involves only tensor contractions and one (truncated) singular value decomposition. Each of these steps have well understood rules under reverse automatic differentiation and are implemented in \textit{TensorKit.jl}.

\bibliography{library_3, library_2, library_1}
\end{document}